\documentclass[aps,prl,twocolumn,showpacs,floatfix,nofootinbib,superscriptaddress,amsmath,amssymb,longbibliography]{revtex4-1}

\usepackage{graphicx}
\usepackage{epsfig}
\usepackage{bm}
\usepackage{color}
\usepackage{float}
\usepackage{dcolumn}
\usepackage{multirow} 
\usepackage{hyperref}

\graphicspath{{.}{../}{figs/}}

\newcommand{\beq}{\begin{equation}}
\newcommand{\eeq}{\end{equation}}
\newcommand{\beqa}{\begin{eqnarray}}
\newcommand{\eeqa}{\end{eqnarray}}

\begin{document}
\title{Structure of $^{78}$Ni from first principles computations}

\author{G.~Hagen}
\affiliation{Physics Division, Oak Ridge National Laboratory,
Oak Ridge, TN 37831, USA} 
\affiliation{Department of Physics and Astronomy, University of Tennessee,
Knoxville, TN 37996, USA} 

\author{G.~R.~Jansen}
\affiliation{National Center for Computational Sciences, Oak Ridge National
Laboratory, Oak Ridge, TN 37831, USA}
\affiliation{Physics Division, Oak Ridge National
Laboratory, Oak Ridge, TN 37831, USA}

\author{T.~Papenbrock}
\affiliation{Physics Division, Oak Ridge National Laboratory,
Oak Ridge, TN 37831, USA} 
\affiliation{Department of Physics and Astronomy, University of Tennessee,
Knoxville, TN 37996, USA} 

\begin{abstract}
Doubly magic nuclei have a simple structure and are the cornerstones
  for entire regions of the nuclear chart. Theoretical insights into
  the supposedly doubly magic $^{78}$Ni and its neighbors are
  challenging because of the extreme neutron-to-proton ratio and the
  proximity of the continuum. We predict the 
   $J^\pi=2_1^+$ state in $^{78}$Ni from a correlation
  with the $J^\pi=2_1^+$ state in $^{48}$Ca using chiral nucleon-nucleon
  and three-nucleon interactions. Our results confirm that $^{78}$Ni
  is doubly magic, and the predicted low-lying states of
  $^{79,80}$Ni open the way for shell-model studies of many more rare
  isotopes.
\end{abstract}

\maketitle 

{\it Introduction} -- Doubly magic nuclei, i.e. nuclei with closed
proton and neutron shells, play a most important role in nuclear
physics~\cite{mayer1955}. They are more strongly bound than their
neighbors, exhibit simple regular patterns, and are the cornerstones
for our understanding of nuclear structure in entire regions of the
Segr{\'e} chart.  In recent years, experiments and theory have made
considerable progress in uncovering the evolution of shell structure
in rare isotopes of
oxygen~\cite{volya2005,hoffman2008,kanungo2009,hoffman2009,otsuka2010,kanungo2011,lunderberg2012,hagen2012a,hergert2013,caesar2013},
calcium~\cite{gallant2012,holt2012,hagen2012b,wienholtz2013,steppenbeck2013,steppenbeck2013b,garciaruiz2016},
and tin~\cite{seweryniak2007,darby2010,jones2010}.

The supposedly doubly magic nucleus $^{78}$Ni (with neutron number 50
and proton number 28) has been the focus of considerable experimental
and theoretical
efforts~\cite{daugas2000,hosmer2005,mazzocchi2005,hakala2008,rajabali2012,sieja2012,tsunoda2014}.
This nucleus is also of astrophysical relevance because it is in the
region of the $r$-process path. Reliable theoretical predictions for
$^{78}$Ni and its neighbors are
challenging~\cite{soma2013,hergert2014}, because of the extreme
neutron-to-proton ratio and the proximity to the neutron dripline. The
large isospin brings to the fore smaller aspects of the nuclear
interaction that are poorly constrained in $\beta$ stable nuclei,
while for weakly bound and unbound nuclear states it is necessary to
include coupling to the particle continuum. We address these
challenges as follows: We employ a set of
interactions~\cite{hebeler2011,ekstrom2015} from chiral effective
field theory
(EFT)~\cite{vankolck1994,epelbaum2009,machleidt2011}. These
interactions consist of nucleon-nucleon ($NN$) and three-nucleon
forces (3NFs)~\cite{epelbaum2002,hebeler2015b}. They reproduce
properties of nuclei with mass numbers $A=2,3,4$ nuclei well, but
differ in binding energies, radii, and spectra of medium-mass
nuclei~\cite{hagen2015}. We include continuum physics by employing the
Berggren basis~\cite{berggren1968,berggren1971,lind1993} which treats
bound-, resonant-, and non-resonant scattering states on equal
footing. The Berggren basis has been extensively used in the
Gamow-shell-model and coupled-cluster computations of weakly bound and
unbound nuclear states, see for example
\cite{michel2002,idbetan2002,hagen2007d}. Finally, using these
ingredients we solve for the structure of $^{78}$Ni and its neighbors
using coupled-cluster theory
~\cite{coester1958,coester1960,cizek1966,kuemmel1978,bishop1991,mihaila2000b,dean2004,kowalski2004,hagen2008,binder2013b},
see Refs.~\cite{bartlett2007,hagen2014} for recent reviews. For the
computation of $J^\pi=2_1^+$excited states in $^{48}$Ca and $^{78}$Ni
we use an implementation of the equation-of-motion (EOM)
coupled-cluster method that properly accounts for
two-particle-two-hole (2$p$-2$h$) excitations.

\begin{figure}[bt]
  \includegraphics[width=1.0\columnwidth]{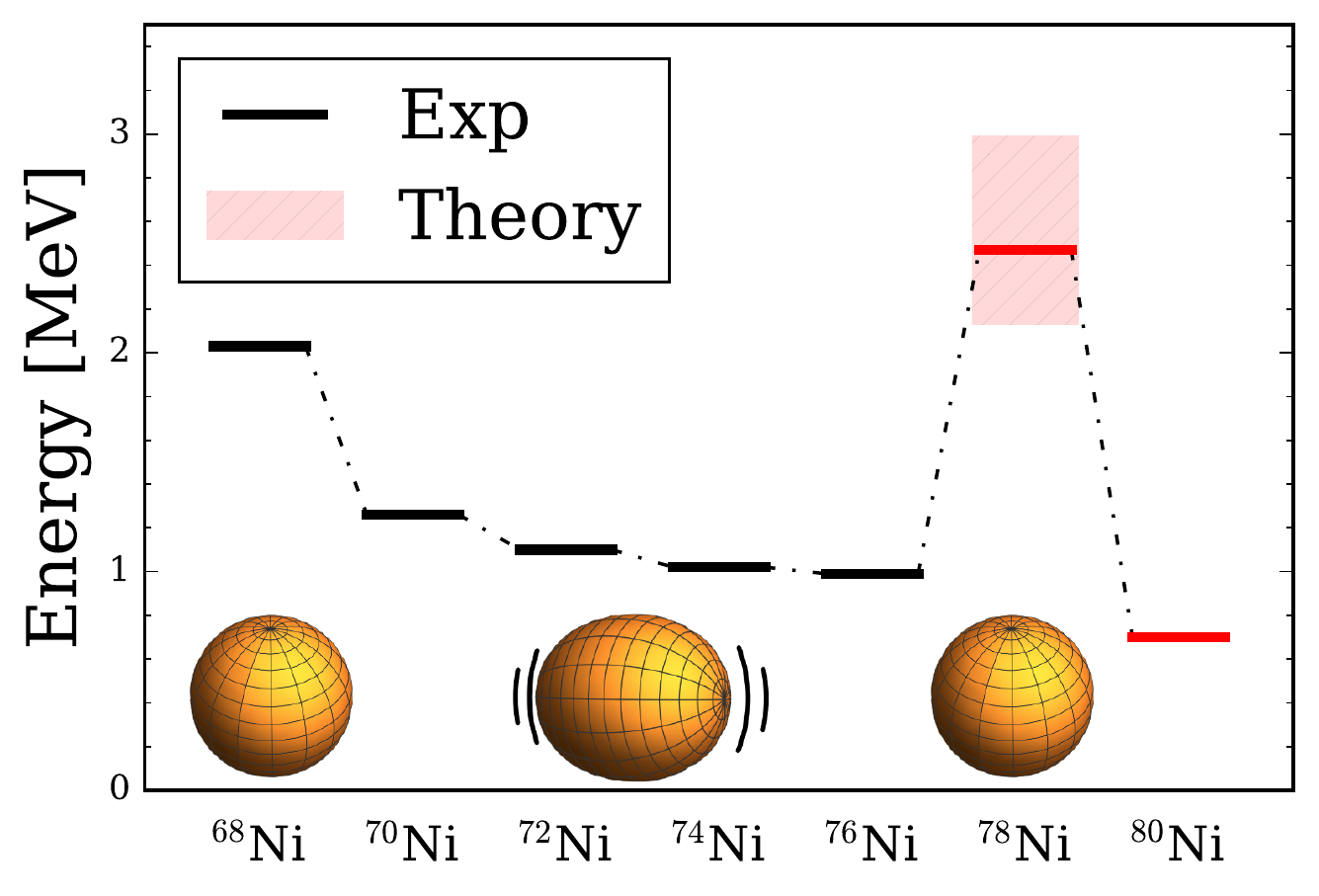}
  \caption{(Color online) Energy of the $2_1^+$ state in neutron-rich
    nickel isotopes for $^{68-76}$Ni from data (black horizontal
    lines) and for $^{78,80}$Ni from first-principles computations
    (red horizontal lines) based on the chiral interaction ``1.8/2.0
    (EM)'' of Ref.~\cite{hebeler2011}. The red shaded area for
    $^{78}$Ni shows the predicted range for the $2_1^+$ state based on
    a correlation between the $2_1^+$ in $^{48}$Ca and $^{78}$Ni using
    a family of chiral interactions (see text and
    Fig.~\ref{Ca48_Ni78_corr} for details).}
  \label{Ni2+evolv}
\end{figure}

As a key indicator of the $^{78}$Ni structure, we focus on the energy
of the first excited $J^\pi=2_1^+$ state. This $2_1^+$ state is at
about 1~MeV of excitation energy in $^{70,72,74,76}$Ni, reflecting a
softness regarding (a collective) quadrupole vibration. In contrast to
these semi magic nuclei, the nucleus $^{68}$Ni exhibits a soft
subshell closure (at neutron number
40)~\cite{sorlin2002,langanke2003}, and its $2_1^+$ state is at about
2~MeV of excitation energy.  This situation is illustrated in
Fig.~\ref{Ni2+evolv}, with experimentally known $2_1^+$ levels shown
as black bars and the computed energies of the $2_1^+$ states in
$^{78,80}$Ni from this Letter.  For $^{78}$Ni the red shaded area
gives the predicted range for the $2_1^+$ state obtained by
correlating relevant observables; details are given below. The
predicted range for the $2_1^+$ state in $^{78}$Ni is considerably
higher than for its neighbors -- indicating that this nucleus is
doubly magic. This is the main result of this Letter. The red bar
marks the result obtained with the interaction ``1.8/2.0(EM)'' from
Ref.~\cite{hebeler2011}, which is singled out because it accurately
reproduces the binding energy of $^{78}$Ni, as well as the nuclei
$^4$He, $^{16}$O, and $^{40,48}$Ca.

This Letter is organized as follows. We briefly summarize the
Hamiltonian and model-spaces that are input to the calculations of
neutron-rich nickel isotopes. We discuss an implementation of
three-particle-three-hole corrections to coupled-cluster computations
of excited states. Using these theoretical ingredients we compute the
first $2_1^+$ state in the doubly magic $^{48}$Ca and in $^{78}$Ni
from a family of chiral $NN$ and 3NFs. From an observed correlation
between the energies of the $2_1^+$ states in $^{48}$Ca and $^{78}$Ni
we obtain a range for the latter. We discuss the relevance of
2$p$-2$h$ excitations in this state. We also give predictions for
other low-lying states in $^{78}$Ni. Finally we focus on the neighbors
of $^{78}$Ni and present predictions for low-lying states in
$^{77,79,80}$Ni.

{\it Hamiltonian and model-space} -- Our coupled-cluster calculations
start from the intrinsic Hamiltonian
\begin{equation}
  \label{intham}
  \hat{H} = \sum_{i<j}\left({({\bf p}_i-{\bf p}_j)^2\over 2mA} + \hat{V}
    _{NN}^{(i,j)}\right) + \sum_{ i<j<k}\hat{V}_{\rm 3N}^{(i,j,k)}.
\end{equation}

We compute the Hamiltonian~(\ref{intham}) using interactions from
Refs.~\cite{hebeler2011,ekstrom2015}. The interactions of
Ref.~\cite{hebeler2011} are based on similarity-renormalization-group
(SRG)~\cite{bogner2007} transformations of $NN$ interactions from
chiral EFT augmented with leading 3NFs from chiral EFT. Here, the
low-energy constants of the 3NFs are adjusted to data from nuclei with
mass numbers $A=3,4$. These interactions yield saturation points for
nuclear matter around the empirical value~\cite{hebeler2011}, and they
yield radii and binding energies in calcium isotopes scattered around
data~\cite{hagen2015}.  The interaction~NNLO$_{\rm sat}$ of
Ref.~\cite{ekstrom2015} by construction yields accurate radii and
binding energies in light nuclei and isotopes of oxygen. It
extrapolates well to calcium isotopes~\cite{hagen2015} and
$^{56}$Ni~\cite{hagen2016}, and within uncertainties reproduces the
empirical saturation point in symmetric nuclear matter.  We employ
these interactions to study systematic sensitivities because a
full-fledged propagation of uncertainties is not yet
possible~\cite{carlsson2016}.  

We use a Hartree-Fock basis constructed from a harmonic oscillator
basis of up to 15 major oscillator shells. To compute weakly bound and
unbound states in $^{79}$Ni we construct a Gamow-Hartree-Fock basis
\cite{michel2004,hagen2007d} by including a Berggren basis for
relevant partial waves and follow Ref.~\cite{hagen2016} for inclusion
of 3NFs. For $^{48}$Ca we use the same model-spaces that were employed
in Ref.~\cite{hagen2015}, while for the neutron-rich nickel isotopes
we perform the calculations at the oscillator frequency $\hbar\omega
=16$~MeV which yields the minimum in energy for the largest
model-space that we consider.  We use the normal-ordered two-body
approximation~\cite{hagen2007a,roth2012,hergert2013b} for the 3NF with
the additional three-body energy cut $E_{3\text{max}}=N_1+N_2+N_3 \leq
16$. Here $N_i = 2n_i + l_i$ refers to the oscillator shell of the
$i^{th}$ particle.

{\it Method } -- We employ the coupled-cluster singles-doubles (CCSD)
approximation in an angular momentum coupled representation in the
computation of the similarity-transformed Hamiltonian $\overline{H}$
(see Refs.~\cite{jansen2012,hagen2014} for details). We include triple
excitations perturbatively using the $\Lambda$-CCSD(T) method
\cite{taube2008} for the computation of the ground-state energy. The
excited $2_1^+$ state is computed with the EOM coupled-cluster method
in the EOM-CCSD \cite{stanton1993} and EOM-CCSD(T)
approximations~\cite{watts1995}. EOM-CCSD has been shown to be
accurate for states that are dominated by 1$p$-1$h$ excitations
\cite{bartlett2007}. In this Letter we go beyond the standard EOM-CCSD
approach and include corrections from 3$p$-3$h$ excitations
perturbatively using the EOM-CCSD(T) approach.  EOM-CCSD(T) capture
the dominant 2$p$-2$h$ excitations in the computation of the $2_1^+$
state in $^{48}$Ca and $^{78}$Ni. This method is a generalization of
the $\Lambda$-CCSD(T) approach for the ground-state energy and
requires the solution of both the left and right EOM-CCSD eigenvalue
problem, and with a non-iterative 3$p$-3$h$ correction computed
perturbatively. We note that the computational cost is considerably
larger than for $\Lambda$-CCSD(T) since we are considering a
non-scalar excitation. In quantum chemistry
applications, EOM-CCSD(T) is an economical and accurate correction
to EOM-CCSD \cite{watson2013}.  Excited states in neighboring nuclei
$^{77,79,80}$Ni are computed as generalized $mp$-$nh$ excited
states~\cite{gour2005,jansen2011,jansen2012} of
$\overline{H}$. Details of this approach are presented in the
review~\cite{hagen2014} and in the supplementary material of
Ref.~\cite{hagen2015}.

{\it Results} -- To probe the quality of the EOM-CCSD(T)
approximation, and for a comparison with data, we also compute the
$2_1^+$ excited state in $^{48}$Ca. For the computation of the $2_1^+$
state in $^{78}$Ni, we employ the same interactions but choose lower
model space frequencies to stabilize the ground-state
energies.

\begin{figure}[tbh]
  \includegraphics[width=1.0\columnwidth]{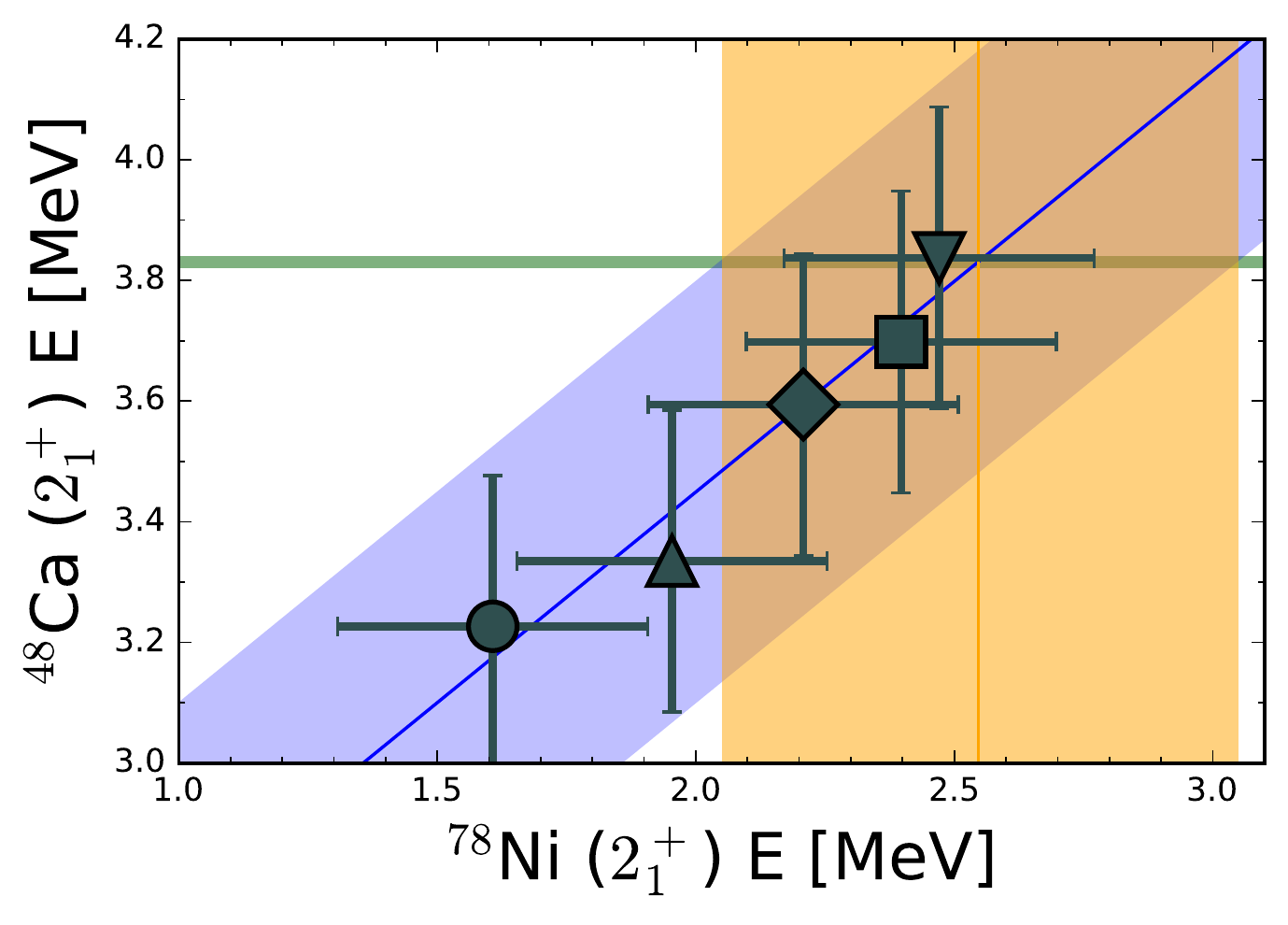}
  \caption{(Color online) Correlation between the energies of the
    $2^+_1$ excited state in $^{48}$Ca and $^{78}$Ni, obtained from
    the interactions NNLO$_{\rm sat}$ (circle), ``2.0/2.0~(PWA)''
    (square), ``2.0/2.0~(EM)'' (diamond), ``2.2/2.0~(EM)'' (triangle
    up), and ``1.8/2.0~(EM)'' (triangle down). The error bars estimate
    uncertainties from enlarging the model space from $N=12$ to
    $N=14$. The thin horizontal line marks the known energy of the
    $2_1^+$ state in $^{48}$Ca. }
  \label{Ca48_Ni78_corr}
\end{figure}

Figure~\ref{Ca48_Ni78_corr} shows that the excitation energy of the
$2_1^+$ state in $^{48}$Ca and $^{78}$Ni are strongly correlated. The
error bars on the individual data points estimate uncertainties from
the method and model-space truncation. We estimate the model-space
uncertainty from enlarging the model space from $N =12$ to $N =14$
which is less than 200~keV for all employed interactions. For the
method we include 10\% of the triples correlation energy as an
uncertainty estimate. We take the average from all interactions and
give a combined uncertainty on the $2_1^+$ state in $^{48}$Ca and
$^{78}$Ni.  A linear fit to the data points, and an encompassing
diagonal uncertainty band is also shown. The thin horizontal line
marks the known energy of the $2_1^+$ state in $^{48}$Ca, and its
intersection with the diagonal band projects out our theoretical
estimate $2.1~{\rm MeV}\lesssim E(2_1^+)\lesssim 3.1$~MeV for the
energy of the $2_1^+$ state in $^{78}$Ni. This band is also shown in
Fig.~\ref{Ni2+evolv}.  We note that two of the five employed
interactions reproduce the energy of the $2_1^+$ state in $^{48}$Ca
within uncertainties. The interaction NNLO$_{\rm sat}$, which
accurately reproduces charge radii in $^{48}$Ca, yields an excitation
energy that is too low. We also note that the origin of the
correlation between the $2_1^+$ states in $^{48}$Ca and $^{78}$Ni
depicted in Fig.~\ref{Ca48_Ni78_corr} is not understood
theoretically. While several such correlations have been reported (and
exploited) in the literature, see, e.g.,
Refs.~\cite{bogner2008,reinhard2010,hagen2015}, only few have been
understood~\cite{platter2005}. The spectroscopy of $^{78}$Ni was
recently measured at RIBF, RIKEN~\cite{taniuchi2016}, and it will be
interesting to compare our theoretical result with data.

\begin{figure}[tbh]
  \includegraphics[width=1.0\columnwidth]{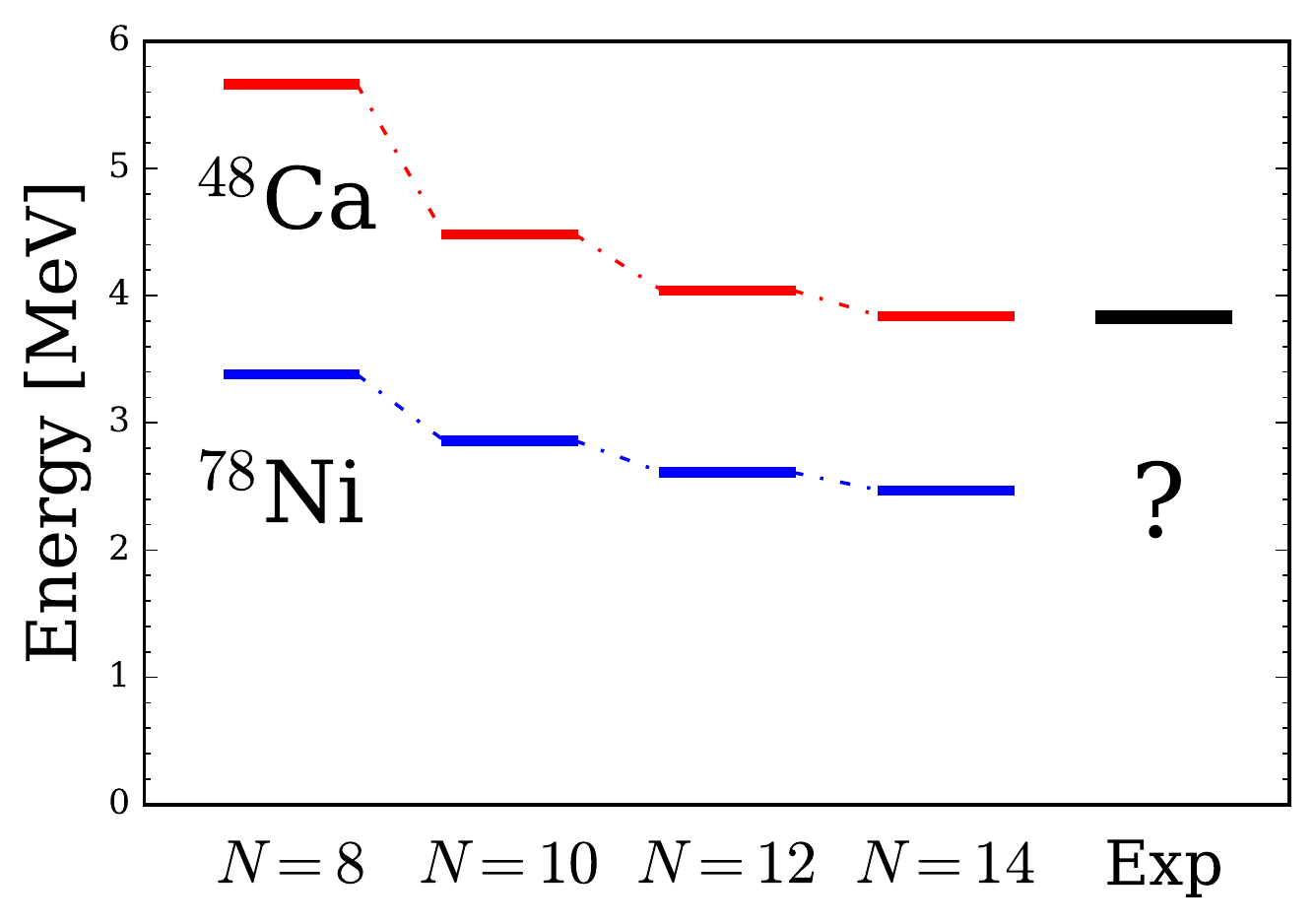}
  \caption{(Color online) Convergence of the first $2_1^+$ excited state
    of $^{48}$Ca and $^{78}$Ni with increasing model-space size and
    compared to data for the interaction ``1.8/2.0 (EM)'' of
    Ref.~\cite{hebeler2011}.}
  \label{Ca48_Ni78_conv}
\end{figure}

For $^{78}$Ni, the convergence of the ground-state energy with respect
to the size of the model space is slow for most of the employed
interactions, and we are only able to achieve convergence for the
softest interaction ``1.8/2.0~(EM)'' of Ref.~\cite{hebeler2011}. For
this interaction the computed binding energy is 637(4)~MeV which
agrees with the value 641~MeV extracted from systematic trends. The
$E_{3\text{max}}$ truncation used for the 3NF is the dominant
uncertainty, and the estimated error of 4~MeV comes from increasing
$E_{3\text{max}}$ from 14 to 16. We note that the convergence is
improved for energy differences. Figure~\ref{Ca48_Ni78_conv} shows the
convergence of the energy of the $2_1^+$ state in $^{48}$Ca and
$^{78}$Ni with increasing size of the model space, obtained for the
interaction ``1.8/2.0~(EM)''. The convergence is qualitatively similar
for the other interactions, and the difference between the $N =12$ and
$N=14$ spaces entered the uncertainties presented in
Fig.~\ref{Ca48_Ni78_corr}.

We note that the interaction ``1.8/2.0~(EM)'' describes of the $2_1^+$
state in $^{48}$Ca and the binding energies for a variety of nuclei
remarkably well.  For example, the computed binding energies for
$^4$He, $^{16}$O and $^{40,48}$Ca are 28.2~MeV, 128~MeV, 348~MeV, and
419~MeV, respectively; they are close to the corresponding
experimental binding energies of 28.2~MeV, 128~MeV, 342~MeV, and
416~MeV.

Let us discuss the effect of 2$p$-2$h$ excitations in the $2_1^+$
excited state of $^{48}$Ca and $^{78}$Ni. Table~\ref{tab1} shows
results for this state using the EOM-CCS, EOM-CCSD and EOM-CCSD(T)
approximations for the interactions used in this work.  We find that
that the inclusion of perturbative 3$p$-3$h$ excitations in
EOM-CCSD(T) reduces the excitation energy by 1-2~MeV for all
interactions when compared to the corresponding EOM-CCSD results. The
triples corrections for the $2_1^+$ state in both $^{48}$Ca and in
$^{78}$Ni amounts to about 20\% of the EOM-CCSD correlation energy
(defined as the difference between the EOM-CCS and EOM-CCSD excitation
energies). We note that the role of 3$p$-3$h$ excitations in the
computation of the $2_1^+$ state in both $^{48}$Ca and in $^{78}$Ni is
considerably larger than the role of 3$p$-3$h$ excitations in the
ground-state. For the ground-state of closed (sub-)shell nuclei the
triples correlation energy typically amounts to about 10\% of the CCSD
correlation energy, see Ref.~\cite{hagen2009b} for an example.

\begin{table}[h]          
\begin{tabular}{|l|r|r|r|r|r|r|}\hline\hline            
\multicolumn{1}{|c|}{} & \multicolumn{3}{c|}{$^{48}$Ca }  & \multicolumn{3}{c|}{$^{78}$Ni } \\ \hline 
 Interaction    & 1$p$-1$h$ & 2$p$-2$h$ & 3$p$-3$h$ & 1$p$-1$h$ & 2$p$-2$h$ & 3$p$-3$h$ \\ \hline  
1.8/2.0 (EM)       & 10.5   &  4.9  &  3.8   &  8.5  &  3.5  &  2.5 \\ 
2.0/2.0 (EM)       & 11.3   &  4.9  &  3.6   &  9.1  &  3.4  &  2.2  \\ 
2.2/2.0 (EM)       & 12.0   &  4.8  &  3.3   &  9.5  &  3.4  &  2.0 \\ 
2.0/2.0 (PWA)      & 12.0   &  5.2  &  3.7   &  9.8  &  3.8  &  2.4  \\ 
  NNLO$_{\rm sat}$ & 14.8   &  5.3  &  3.2   & 12.2  &  3.8  &  1.6  \\\hline\hline        
\end{tabular}        
\caption{Results for the excitation energy (in MeV) of the $2_1^+$
  state in $^{48}$Ca and $^{78}$Ni computed in the EOM-CCS (denoted by
  1$p$-1$h$), EOM-CCSD (denoted by 2$p$-2$h$) and EOM-CCSD(T) (denoted
  by 3$p$-3$h$) approximations. The interactions labeled (EM) and
  (PWA) are taken from Ref.~\cite{hebeler2011} and NNLO$_{\rm sat}$ is
  from Ref.~\cite{ekstrom2015}.}
\label{tab1}  
\end{table}

Our analysis shows that 2$p$-2$h$ excitations are significant for the
$2_1^+$ state in $^{48}$Ca and $^{78}$Ni, and that a precise
description of this state therefore requires EOM-CCSD(T). This finding
is somewhat surprising, because the collective $2_1^+$ state is
usually thought of as a coherent superposition of 1$p$-1$h$
excitations~\cite{ringschuck}. However, a simple shell-model argument
suggests that 2$p$-2$h$ excitations should yield significant
corrections. In the doubly-magic $^{48}$Ca for instance, no 1$p$-1$h$
excitations of protons near the Fermi surface can generate a $2^+$
state, as one need at least 2$p$-2$h$ excitations from the $sd$ shell
to the $pf$ shell to yield a $2^+$ state. Following the same
reasoning, a computation of the electric quadrupole transition in
$^{48}$Ca will have significant 2$p$-2$h$ contributions since this
observable measures mostly the excitations of protons.  Similarly, we
find that for $^{78}$Ni 2$p$-2$h$ excitations of neutrons near the
Fermi surface have significant contributions to the low-lying $2_1^+$
state. In the naive shell-model picture the $g_{9/2}$ orbital is the
last filled neutron shell with $s_{1/2},d_{5/2}, d_{3/2}, g_{7/2}$
shells being the next unoccupied orbitals closest to the Fermi
surface.  A $2^+$ state near the Fermi surface can be generated via
1$p$-1$h$ excitations of neutrons from the $g_{9/2}$ to the $d_{5/2},
g_{7/2}$ orbitals, but 2$p$-2$h$ excitations are necessary to utilize
the low-lying $s_{1/2}$ and $d_{3/2}$ orbitals.  As shown in
Tab.~\ref{tab1} the effect of 2$p$-2$h$ excitations from the $g_{9/2}$
to the $s_{1/2}$ and $d_{3/2}$ orbitals is significant in the $2^+$
state of $^{78}$Ni. As we will see below the ${1/2}^+$ state is
actually the lowest state in $^{79}$Ni.

\begin{figure}[t]
  \includegraphics[width=1.0\columnwidth]{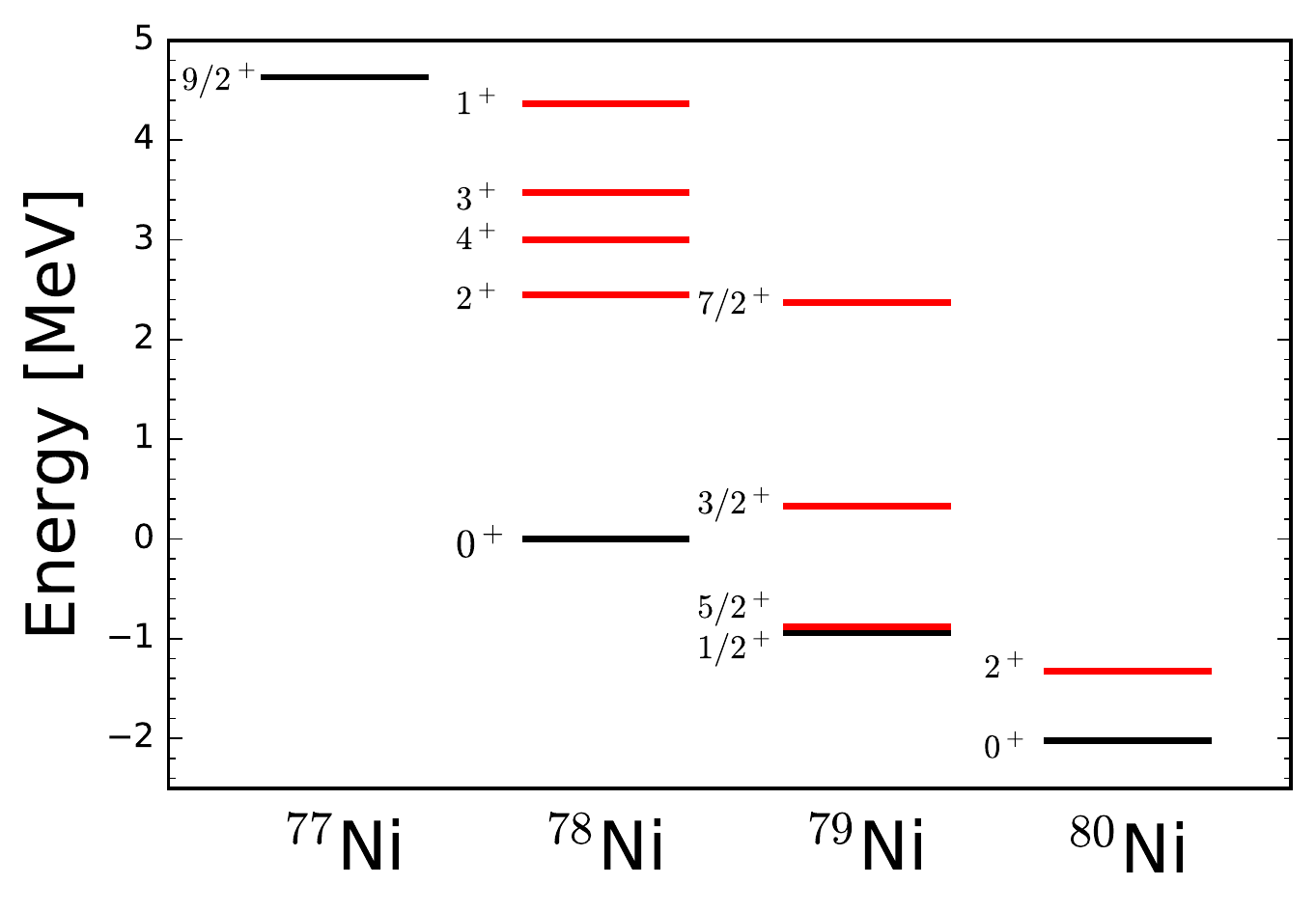}
  \caption{(Color online) Low-lying states in $^{77-80}$Ni with
    respect to the ground-state of $^{78}$Ni computed with the
    interaction ``1.8/2.0 (EM)'' of Ref.~\cite{hebeler2011}. The
    ground states are shown in black, while excited states are shown
    in red.}
  \label{Ni77_80_spectra}
\end{figure}

Shell closures manifest themselves in several observables. Besides the
energy of the $2_1^+$ state, separation energies also yield valuable
information. For the computation of other low-lying states in
$^{78}$Ni and in the neighboring nuclei $^{77,79,80}$Ni, we limit
ourselves to the ``1.8/2.0~(EM)'' interaction because this interaction
yields converged results with respect to the model space and accurate
binding energies from $^4$He to $^{16}$O to $^{48}$Ca to
$^{78}$Ni. For $^{79}$Ni, we employed a Berggren basis for the
$s_{1/2}$, $d_{5/2}$ and $d_{3/2}$ partial waves because of the
proximity of the continuum. For the $g_{7/2}$ partial wave we use the
harmonic-oscillator basis, because the large centrifugal barrier
reduces the impact of the coupling to the continuum. The resulting
spectra are shown in Fig.~\ref{Ni77_80_spectra} relative to the
ground-state energy of $^{78}$Ni. For $^{78}$Ni we predict low-lying
$1_1^+,3_1^+,4_1^+$ excited states all below the neutron-emission
threshold. The ratio of the excited $4_1^+$ state with the
$2_1^+$ state is 1.2, which is consistent with $^{78}$Ni being a
doubly magic nucleus. Due to the high computational cost the
$1_1^+,3_1^+,4_1^+$ excited states in $^{78}$Ni were computed with $N
= 12$; the triples correlation energy for the $4_1^+$ state was
well converged for $N=10$.  The theoretical result for the
neutron-separation energies in $^{78,79}$Ni are $S_n\approx 4.5$~MeV
and $S_n\approx 1$~MeV, respectively, which are consistent with
5450(950)~keV and 1650(1130)~keV from systematics~\cite{wang2012}. For
$^{79}$Ni we find that the inclusion of the continuum impacts the
level ordering and lowers the ${1/2}^+$ state by about 1~MeV, the
${5/2}^+$ state by about 0.5~MeV, and the unbound ${3/2}^+$ state by
about 0.7~MeV, as compared to a calculation using harmonic oscillator
functions only.

The ${1/2}^+$ ground-state of $^{79}$Ni is quasi-degenerate with the
${5/2}^+$ state. This finding mirrors the results of
Refs.~\cite{hagen2012b,hagen2013,hagen2016}, where the inclusion of
continuum effects also impacted the energies and level ordering of
unbound states in the neutron-rich calcium isotopes $^{53,55,61}$Ca.
The ground-state of $^{80}$Ni is bound by 2~MeV with respect to
$^{78}$Ni, thereby setting the neutron dripline beyond $^{80}$Ni. This
is consistent with mean-field surveys~\cite{erler2012}. The
two-neutron separation $S_{2n}(^{80}{\rm Ni})\approx 2$~MeV is
significantly smaller than the estimate $S_{2n}(^{78}{\rm
  Ni})=$~8660(950)~keV~\cite{wang2012} -- consistent with expectations
for a doubly magic nucleus. The $2_1^+$ state in $^{80}$Ni is computed
to be 0.7~MeV above its ground state. The combined results of this
study -- a relatively high-lying $2_1^+$ state in $^{78}$Ni, the
marked difference of neutron-separation energies between $^{79}$Ni and
$^{78}$Ni, and of two-neutron separation energies between $^{80}$Ni
and $^{78}$Ni, respectively, indicate the strength of the shell
closure at neutron number 50.

{\it Conclusions} -- We presented first-principles computations of the
structure of $^{78}$Ni and its neighbors. Correlating the $2_1^+$
energies in $^{78}$Ni and $^{48}$Ca leads to the prediction $2.1~{\rm
  MeV}\lesssim E(2_1^+)\lesssim 3.1$~MeV for the energy of the $2_1^+$
state in $^{78}$Ni. Neutron separation energies and two-neutron
separation energies confirm the picture of the shell closure at
neutron number 50, and the theoretical results put the neutron
dripline beyond $^{80}$Ni. We also made predictions for low-lying
states in $^{77,78,79,80}$Ni that can be confronted by experiment. As
a useful theoretical tool, a relatively soft chiral interaction
emerged as being in good agreement with binding energies and low-lying
excitations from $^4$He, to $^{16}$O, to $^{40,48}$Ca to
$^{78}$Ni. This study paves the way to theoretical predictions in
heavy rare isotopes.

\begin{acknowledgments}
  We thank Kai Hebeler for providing us with matrix elements in Jacobi
  coordinates for the three-nucleon interaction at
  next-to-next-to-leading order. This work was supported by the Office
  of Nuclear Physics, U.S. Department of Energy, under grants
  DE-FG02-96ER40963, DE-SC0008499 (NUCLEI SciDAC collaboration), and
  the Field Work Proposal ERKBP57 at Oak Ridge National Laboratory
  (ORNL).  Computer time was provided by the Innovative and Novel
  Computational Impact on Theory and Experiment (INCITE) program. This
  research used resources of the Oak Ridge Leadership Computing
  Facility located at ORNL, which is supported by the Office of
  Science of the Department of Energy under Contract No.
  DE-AC05-00OR22725.
\end{acknowledgments}

\bibliography{refs}

\end{document}